\documentclass[traditabstract]{aa} 
\usepackage{graphicx}
\usepackage{txfonts}
\newcommand\kms{{\rm\,km\,s^{-1}}}

\begin{document}

\title{Massive runaway stars in the Small Magellanic Cloud}

\author{V.V.~Gvaramadze\inst{1,2,3}
\and J.~Pflamm-Altenburg\inst{1} \and P.~Kroupa\inst{1}}

\institute{Argelander-Institut f\"{u}r Astronomie, Universit\"{a}t
Bonn, Auf dem H\"{u}gel 71, 53121 Bonn,
Germany\\\email{jpflamm@astro.uni-bonn.de (JP-A);
pavel@astro.uni-bonn.de (PK)} \and Sternberg Astronomical
Institute, Moscow State University, Universitetskij Pr. 13, Moscow
119992, Russia\\\email{vgvaram@mx.iki.rssi.ru (VVG)} \and Isaac
Newton Institute of Chile, Moscow Branch,
Universitetskij Pr. 13, Moscow 119992, Russia\\
}

\date{Received 27 August 2010/ Accepted 11 October 2010}

\abstract{Using archival {\it Spitzer Space Telescope} data, we
identified for the first time a dozen runaway OB stars in the
Small Magellanic Cloud (SMC) through the detection of their bow
shocks. The geometry of detected bow shocks allows us to infer the
direction of motion of the associated stars and to determine their
possible parent clusters and associations. One of the identified
runaway stars, AzV\,471, was already known as a high-velocity star
on the basis of its high peculiar radial velocity, which is offset
by $\simeq 40 \, \kms$ from the local systemic velocity. We
discuss implications of our findings for the problem of the origin
of field OB stars. Several of the bow shock-producing stars are
found in the confines of associations, suggesting that these may
be ``alien" stars contributing to the age spread observed for some
young stellar systems. We also report the discovery of a
kidney-shaped nebula attached to the early WN-type star SMC-WR3
(AzV\,60a). We interpreted this nebula as an interstellar
structure created owing to the interaction between the stellar
wind and the ambient interstellar medium.}


\maketitle


\section{Introduction}
%

Gravitational interaction between massive stars in the cores of
young star clusters results in ejection of a significant fraction
of these stars from the parent clusters (e.g. Pflamm-Altenburg \&
Kroupa \cite{pf06};  Moeckel \& Bate \cite{mo10}). The ejected
(runaway) stars form the population of field OB stars that end
their lives in supernova explosions hundreds of parsecs from their
birthplaces. The high velocities of runaway OB stars can be
revealed either directly, via measurement of their proper motions
and/or radial velocities (e.g. Moffat et al. \cite{mo98};
Mdzinarishvili \& Chargeishvili \cite{md05}; Massey et al.
\cite{ma05}; Evans et al. \cite{ev10}), or indirectly, through the
detection of bow shocks generated ahead of supersonically moving
stars (Gvaramadze \& Bomans \cite{gv08b}; Gvaramadze, Kroupa \&
Pflamm-Altenburg \cite{gv10a}, hereafter Paper\,I). The proper
motions can be used to trace the trajectories of the field stars
back to their parent clusters, hence to prove their runaway nature
(e.g. Hoogerwerf, de Bruijne \& de Zeeuw \cite{ho01}; Schilbach \&
R\"{o}ser \cite{sc08}). But the proper motions can be measured
with a high confidence only for very high-velocity or relatively
nearby stars. Measurements of radial velocities can be used to
prove the runaway nature of the field stars as well, but taken
alone they do not allow us to infer the birth clusters of the
stars. For distant OB stars, detection of bow shocks provides a
unique possibility not only to show that these stars are runaways,
but also to infer the direction of their motion on the sky and
thereby to search for their likely parent clusters and OB
associations (Gvaramadze \& Bomans \cite{gv08b}; Paper\,I).

The necessary conditions for the existence and detection of bow
shocks are that the associated stars are moving through the
interstellar medium with a supersonic velocity ($> 10 \, \kms$, in
the warm ionized medium) and that the ambient medium is dense
enough to ensure a sufficiently high emission measure of the bow
shocks (either in H$_\alpha$ or in the infrared). Observations
show that $\la 20$ per cent of runaway OB stars are associated
with (detectable) bow shocks (Van Buren, Noriega-Crespo, \& Dgani
\cite{va95}; Noriega-Crespo, Van Buren \& Dgani \cite{no97};
Huthoff \& Kaper \cite{hu02}).

Identification of runaway stars via the detection of their bow
shocks was used by us to search for OB stars running away from
young star clusters in the Milky Way (Gvaramadze \& Bomans
\cite{gv08a},b; Gvaramadze et al. \cite{gv10b}) and to prove the
runaway nature of the very massive field stars in the Large
Magellamic Cloud (LMC; Paper\,I). In the latter case, we detected
bow shocks around several field OB stars (the first-ever detection
of bow shocks generated by extragalactic stars), one of which, the
O2\,V((f*)) star \object{BI\,237}, had been earlier proposed as a
candidate runaway star owing to its high peculiar radial velocity
(Massey et al. \cite{ma05}). A logical extension of this work is
to use the same approach to identify runaway stars in the Small
Magellanic Cloud (SMC), after the LMC our next closest dwarf
galaxy, where bow shocks can still be resolved with modern
infrared telescopes. In this paper, we present the results of a
systematic search for bow shock-producing stars in the SMC, using
the {\it Spitzer Space Telescope} archival data. Throughout the
paper we use a distance of $\simeq 60$ kpc for the SMC (Hilditch,
Howarth \& Harries \cite{hi05}) so that $1\arcmin$ corresponds to
$\simeq 17$ pc.

\section{Search for bow shocks in the SMC}

\begin{table*}
  \caption{Summary of bow shock-producing stars in the SMC.}
  \label{tab:phot}
  \renewcommand{\footnoterule}{}
\begin{tabular}{lcccccccc}
\hline \hline
Star & RA  & Dec. & Spectral & Ref. & Association & Age & Separation \\
 & (J2000) & (J2000) & type & & or cluster & (Myr) & (pc)  \\
\hline
\object{[MB2000]\,37} &  00 47 30.12 & $-$73 05 07.5 & B9\,Ia & 1 & \object{DEM S 47} & 8$^{a}$ & 210 \\ 
\object{AzV\,26} & 00 47 50.01 & $-$73 08 21.0 & O6\,I(f) & 2 & \object{SMC ASS 13} & $\leq 4^{a}$ & 230 \\ 
\object{[M2002]\,SMC\,9824} &  00 48 02.63 & $-$73 16 38.7 & OB & 3 & \object{DEM S 45} & 10$^{a}$ & 160 \\ 
\object{[M2002]\,SMC\,12977} & 00 49 14.13 & $-$73 14 42.6 & OB & 3 & \object{DEM S 32} & 10$^{a}$ & 160 \\ 
\object{AzV\,56} & 00 49 51.26 & $-$72 55 45.3 & B2.5\,Ia & 4 & \object{DEM S 55} & 6$^{a}$ & 150 \\ 
\object{AzV\,75} & 00 50 32.39 & $-$72 52 36.5 & O5.5\,I(f) & 5 & \object{SMC ASS 22} & $\leq 4^{a}$ & 100 \\ 
\object{AzV\,77} & 00 50 33.54 & $-$72 47 45.0 & O7\,III & 6 & \object{DEM S 54} & 8$^{a}$ & 90 \\ 
\object{AzV\,242} & 01 00 06.88 & $-$72 13 57.5 & B1\,Ia & 4 & \object{DEM S 118} & 8$^{a}$ & 290 \\ 
\object{AzV\,398} & 01 06 09.81 & $-$71 56 00.8 & O8.5\,If & 7 & \object{NGC 371} & 5$^{a}$ & 270 \\ 
\object{AzV\,429} & 01 07 59.85 & $-$72 00 53.9 & O7\,V & 6 & \object{[BS95] 134} & -- & 150 \\ 
\object{AzV\,471} & 01 13 00.41 & $-$73 17 04.1 & B0\,(III) & 8 & \object{NGC\,456} & 10$^{(b)}$ & 60 \\ 
\object{[M2002]\,SMC\,83962} & 01 31 54.97 & $-$73 27 23.4 & OB & 3 & \object{NGC\,602c} & 3$^{(c)}$ & 70 \\ 
\hline
\end{tabular}
\tablefoot{
 \tablefoottext{a}{Chiosi et al. (\cite{ch06}).}
 \tablefoottext{b}{Hodge (\cite{ho83}).}
 \tablefoottext{c}{Massey, Waterhouse \& DeGioia-Eastwood (\cite{ma00}).}
 }
 \tablebib{(1)~Oblak \& Chareton \cite{ob81}; (2)~Massey et al. \cite{ma04}; (3)~Oey, King \& Parker \cite{oe04}; (4)~Lennon \cite{le97};
 (5)~Massey et al. \cite{ma09};  (6)~Garmany, Conti \& Massey \cite{ga87}; (7)~Massey \& Duffy \cite{ma01}; (8)~Evans et al. \cite{ev04}.}
 \label{tab:run}
\end{table*}
\begin{figure}
\includegraphics[width=9cm]{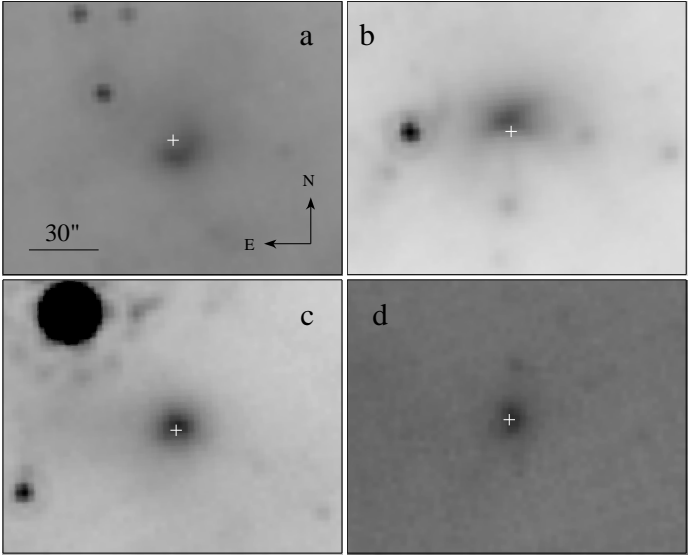}
\centering \caption{MIPS 24$\mu$m images of four bow shocks
associated with (a) [MB2000]\,37, (b) AzV\,26, (c) AzV\,77, and
(d) AzV\,471. The positions of the stars are marked by crosses.
The orientation and the scale of the images are the same.}
\label{fig:bows}
\end{figure}
\begin{figure}
\includegraphics[width=9cm]{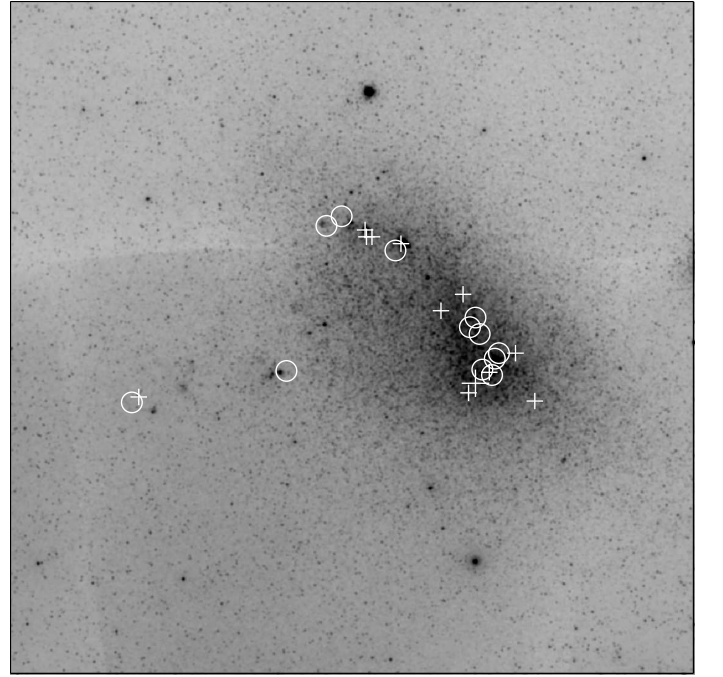}
\centering \caption{$6\degr \times 6\degr$ DSS-II image of the SMC
and its environments, with the positions of bow shock-producing
stars marked by circles. Crosses indicate the positions of all 12
known Wolf-Rayet stars in the SMC, one of which, SMC-WR3, is
marked out by a large cross (see text for details). North is up
and east to the left.} \label{fig:SMCbows}
\end{figure}

To search for bow shocks in the SMC, we utilized the publicly
available imaging data from the {\it Spitzer Space Telescope}
archive. Using the Leopard software, we retrieved images of the
SMC obtained with the Multiband Imaging Photometer for {\it
Spitzer} (MIPS; Rieke et al. \cite{ri04}) in the framework of the
S$^3$MC (ID 3316\footnote{http://celestial.berkeley.edu/spitzer/};
Bolatto et al. \cite{bo07}) and the SAGE-SMC (ID
40245\footnote{http://sage.stsci.edu/}) projects. Our search was
directed to the detection of arclike structures associated with
known OB stars. The typical linear size of bow shocks produced by
runaway OB stars is several parsec, which at the distance of the
SMC corresponds to the angular size of several tens of arc
seconds; we therefore limited ourselves to the search for
structures of this angular size. From our experience in the search
for bow shocks generated by runaway OB stars (e.g. Paper\,I), we
know that the bow shocks are visible mostly in MIPS $24\,\mu$m
images, so that we utilized the $24\,\mu$m data alone. The angular
resolution of these data of 6 arc seconds is comparable to or
several times smaller than the expected angular size of bow shocks
in the SMC.

Visual inspection of the MIPS $24\,\mu$m images of the SMC
revealed numerous diffuse structures whose arc-like morphology
suggests that they could be bow shocks. Using the SIMBAD
database\footnote{http://simbad.u-strasbg.fr/simbad/} and the
VizieR catalogue access
tool\footnote{http://webviz.u-strasbg.fr/viz-bin/VizieR}, we found
that most of these structures (see Fig.\,\ref{fig:bows} for four
examples) are associated with known OB stars, which strengthens
their interpretation as bow shocks. The details of the bow
shock-producing stars (listed in order of their RA) are presented
in Table\,\ref{tab:run}. The equatorial coordinates of the stars
were taken from the 2MASS catalogue (Skrutskie et al.
\cite{sk06}). Column 5 gives the references from which the
spectral classification was obtained. Note that the spectral type
of [MB2000]\,37 was estimated photometrically (Oblak \& Chareton
\cite{ob81}). Similarly, three other stars in the table were
identified by Oey et al. (\cite{oe04}) as OB stars on the basis of
the UBVR survey data of Massey (\cite{ma02}) for the SMC.
Follow-up spectroscopy of these four stars is required to refine
their spectral types and thereby to check whether their ages are
consistent with the ages of their likely parent clusters and
associations (see below).

Figure\,\ref{fig:SMCbows} shows the optical (red band) image of
the SMC from the Digitized Sky Survey II (DSS-II; McLean et al.
\cite{mc00}) with the positions of bow shock-producing stars
indicated by circles\footnote{The image was generated by the
NASA's {\it SkyView} facility (McGlynn, Scollick \& White
\cite{mc96}).}. As expected, the majority of bow shock-producing
stars reside in the main body (the bar) of the SMC, where most of
star-forming regions and young star clusters are located (e.g.,
Harris \& Zaritsky \cite{ha04}). For illustrative purposes, we
indicated in Fig.\,\ref{fig:SMCbows} the positions of all 12 known
Wolf-Rayet (WR) stars in the SMC, which represent the population
of very massive and very young stars (Massey et al. \cite{ma00})
and therefore closely trace the regions of ongoing star formation.
Two of the bow shock-producing stars, namely \object{AzV\,471} and
[M2002]\,SMC\,83962, were detected in the wing of the SMC (the
brightest section of a stellar system extending eastward from the
SMC towards the LMC).

\begin{figure}
\includegraphics[width=9cm]{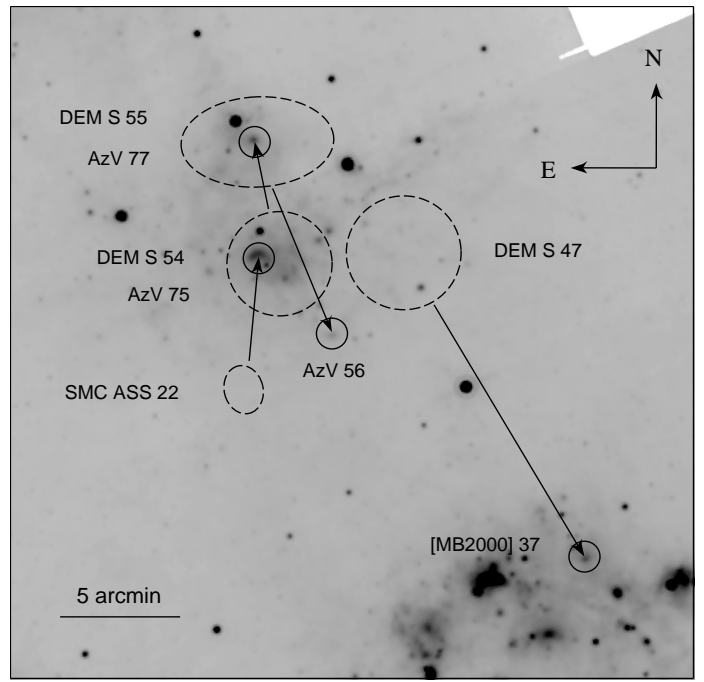}
\centering \caption{MIPS $24\,\mu$m image of the environment of
four bow shock-producing stars (indicated by solid circles). The
arrows show the direction of motion of the stars, as suggested by
the geometry of their bow shocks. The approximate boundaries of
the possible birth associations are shown by dashed circles and
ellipses (see text for details). At the distance of the SMC,
$1\arcmin$ corresponds to $\simeq 17$ pc.} \label{fig:1567}
\end{figure}
\begin{figure}
\includegraphics[width=9cm]{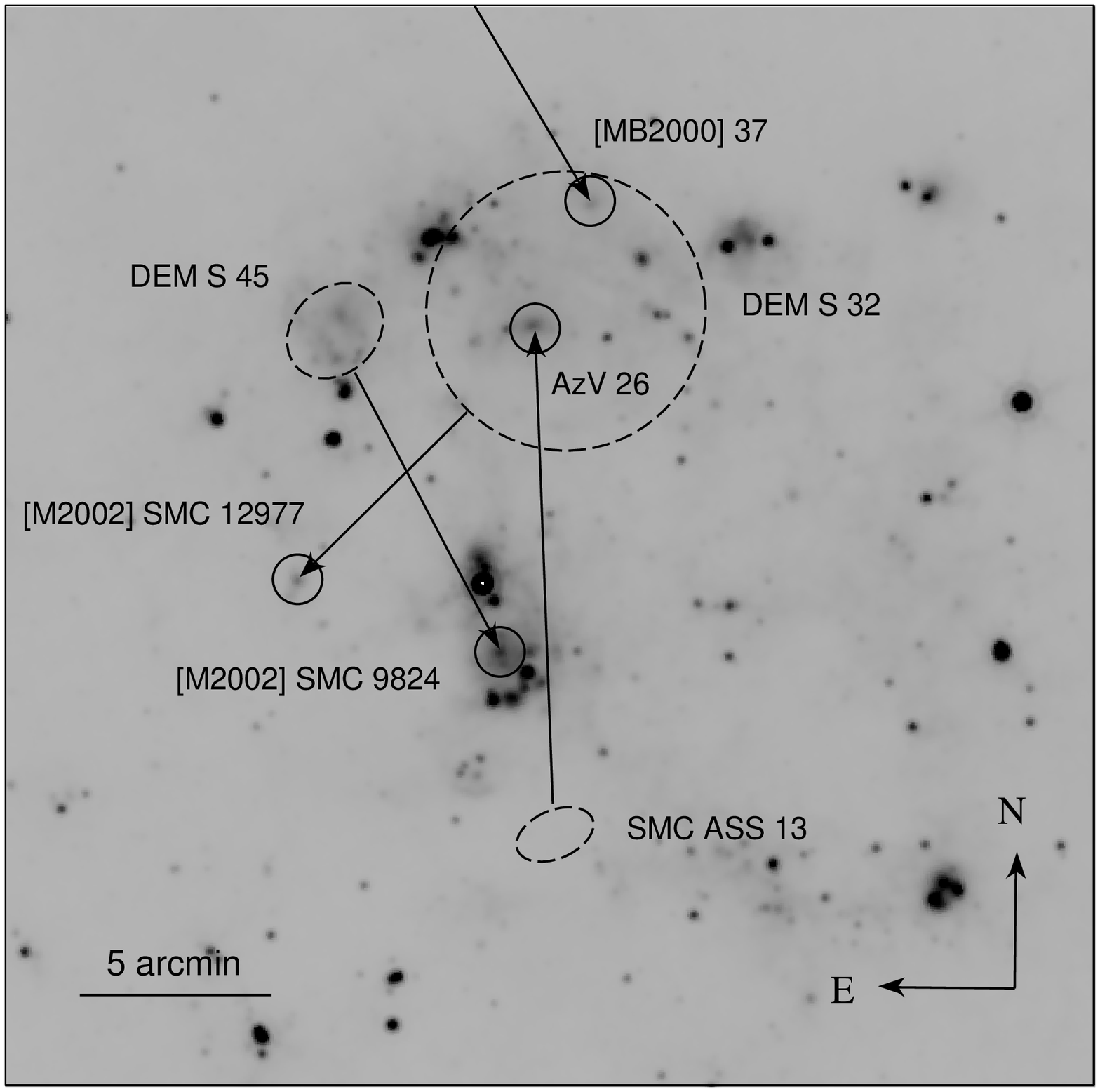}
\centering \caption{MIPS $24\,\mu$m image of the environment of
four bow shock-producing stars (indicated by solid circles). The
arrows show the direction of motion of the stars, as suggested by
the geometry of their bow shocks. The approximate boundaries of
the possible birth associations are shown by dashed circles and
ellipses.} \label{fig:234}
\end{figure}
\begin{figure}
\includegraphics[width=9cm]{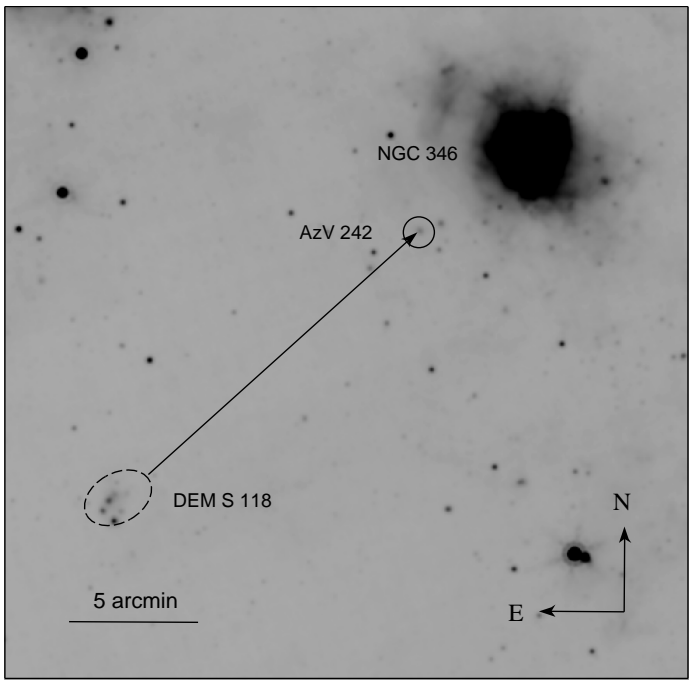}
\centering \caption{MIPS $24\,\mu$m image of the area to the
southeast of the giant star-forming region NGC\,346 with the bow
shock-producing star AzV\,242 (indicated by a solid circle) and
its possible parent association DEM\,S\,118 (shown by a dashed
ellipse). The arrow shows the direction of motion of the star, as
suggested by the geometry of the bow shock.} \label{fig:8}
\end{figure}

To identify possible birthplaces of the bow shock-producing stars,
we searched for nearest young ($\la 10$ Myr) star clusters and
associations, located in the opposite direction of stellar motion
(inferred from the geometry of the bow shocks). The coordinates of
clusters and associations and their approximate boundaries were
taken from the catalogue by Bica \& Dutra (\cite{bi00}), while the
ages were taken from Chiosi et al. (\cite{ch06}), Massey et al.
(\cite{ma00}), and Hodge (\cite{ho83}). For one of the
associations, [BS95]\,134, we were unable to find the age
estimate. The results from this search are summarized in
Table\,\ref{tab:run}, where we give also the separations (in
projection) of the stars from the centres of their possible parent
clusters and associations, and the ages of the clusters and the
associations. Using the figures from the last two columns of
Table\,\ref{tab:run} and assuming that the stars were ejected in
the field at the very beginning of dynamical evolution of the
parent stellar systems, one can estimate their minimum ejection
velocities, i.e., ejection velocity ($\kms$) = separation (pc)/age
(Myr). For most stars, the (minimum) ejection velocities are quite
moderate ($< 30 \, \kms$), so that they cannot be formally
classified as runaways (Blaauw \cite{bl61}). On the other hand,
one cannot exclude the possibility that the detected bow
shock-producing stars were ejected from the more distant clusters
and associations or that they left their birthplaces after several
Myr of cluster evolution (e.g. because of the binary supernova
explosions). High-precission proper motion measurements (e.g. with
the future space astrometry mission {\it Gaia}) are therefore
required to resolve the issue.

Figure\,\ref{fig:1567} shows four bow shocks (indicated by solid
circles) and the trajectories of their associated stars (shown by
arrows), as suggested by the geometry of the bow shocks. The
origin of the trajectories corresponds to possible birth
associations, whose boundaries are indicated by dashed circles and
ellipses. One can see that two stars, AzV\,75 and AzV\,77, are
located (at least in projection) within the associations
DEM\,S\,54 and DEM\,S\,55, respectively. The detection of the bow
shocks generated by these stars (see Fig.\,\ref{fig:bows}c for the
MIPS 24 $\mu$m image of the bow shock generated by AzV\,77)
implies that they are runaways and that they therefore were
injected into their host associations from the nearby ones,
probably from the associations SMC\,ASS\,22 and DEM\,S\,54 (cf.
Gvaramadze \& Bomans \cite{gv08b}; Paper\,I). The geometry of the
bow shock-producing stars, AzV\,56 and [MB2000]\,37
(Fig.\,\ref{fig:bows}a), suggests that they were ejected from
DEM\,S\,55 and DEM\,S\,47, respectively.

Figure\,\ref{fig:234} shows three other bow shocks and the bow
shock (generated by [MB2000]\,37) already shown in
Fig.\,\ref{fig:1567}. The orientation of the bow shocks around
[MB2000]\,37 (Fig.\,\ref{fig:bows}a), AzV\,26
(Fig.\,\ref{fig:bows}b) and [M2002]\,SMC\,12977 suggests that the
first two stars were injected into the association DEM\,S\,32,
while the third star, on the contrary, was ejected from the
association. The orientation of the bow shock associated with
[M2002]\,SMC\,9824 is consistent with the possibility that this
star was ejected from DEM\,S\,45.

Figure\,\ref{fig:8} shows the area to the southeast of the giant
star-forming region NGC\,346. The rich population of very massive
stars (including the Luminous Blue Variable/WR binary HD\,5980;
e.g. Koenigsberger et al. \cite{ko10}) and significant mass
segregation in NGC\,346 (e.g., Hennekemper et al. \cite{he08};
Sabbi et al. \cite{sa08}) suggest that this region should be
effective in producing runaway stars. It might therefore be
expected that some of the ejected stars would manifest themselves
in bow shocks. Despite this expectation, we found only one bow
shock-producing star around NGC\,346, and this star, AzV\,242, is
instead moving towards NGC\,346.

\begin{figure}
\includegraphics[width=9cm]{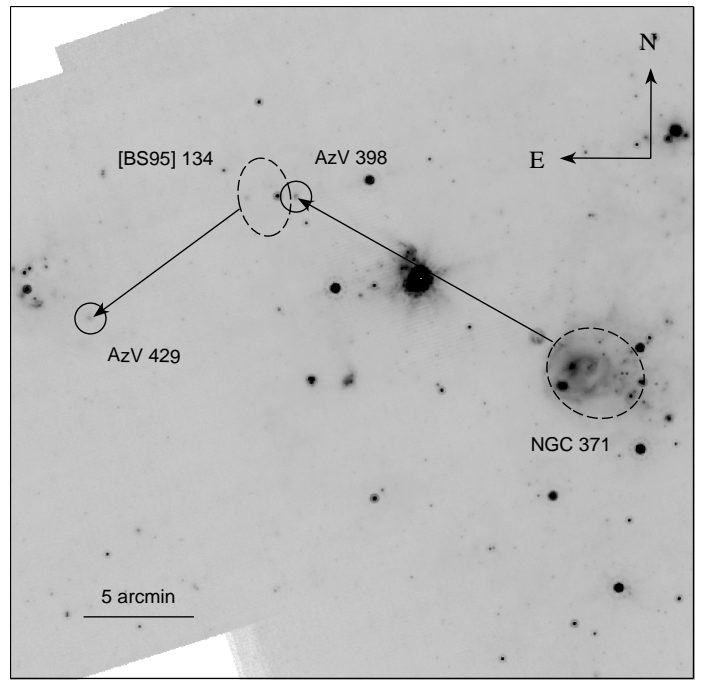}
\centering \caption{MIPS $24\,\mu$m image of the environment of
AzV\,398 and AzV\,429 and their bow shocks (indicated by solid
circles). The arrows show the direction of motion of the stars, as
suggested by the geometry of their bow shocks. The approximate
boundaries of the possible birth cluster NGC\,371 and the
association [BS95] 134 are shown by dashed ellipses. A small
ring-like structure near the northeast edge of NGC\,371 is the
supernova remnant 1E\,0102.2$-$7219.} \label{fig:910}
\end{figure}

\begin{figure}
\includegraphics[width=9cm]{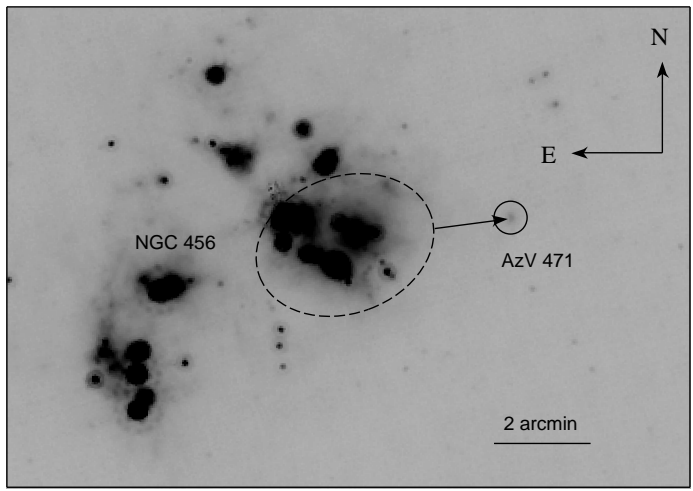}
\centering \caption{MIPS $24\,\mu$m image of the star-forming
region N83/N84 with the association NGC\,456 (indicated by a
dashed ellipse) and the bow shock generated by the BO\,(III) star
AzV\,471 (shown by a circle).} \label{fig:11}
\end{figure}

\begin{figure}
\includegraphics[width=9cm]{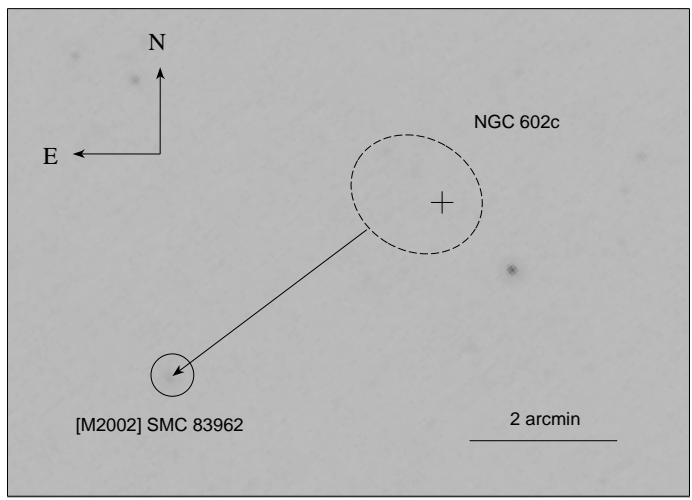}
\centering \caption{MIPS $24\,\mu$m image of the association
NGC\,602c (indicated by a dashed ellipse) and the bow shock
generated by the OB star [M2002]\,SMC\,83962 (shown by a circle).
The cross indicates the position of the WO4+O\,V star SK\,188.}
\label{fig:12}
\end{figure}

Figures\,\ref{fig:910}, \ref{fig:11}, and \ref{fig:12} show the
remaining four bow shock-producing stars detected in the SMC. Two
of these stars, AzV\,398 and AzV\,429, are located near the
northeastern edge of the SMC and were probably ejected from the
associations, NGC\,371 and [BS95]\,134, respectively
(Fig.\,\ref{fig:910}). The bow shock-producing star AzV\,471 is
located not far from the N83/N84 star-forming region (Henize
\cite{he56}) of the inner wing of the SMC (Fig.\,\ref{fig:11}).
The orientation of the bow shock (see Fig.\,\ref{fig:bows}d) is
consistent with the possibility that the star was ejected from the
association NGC\,456. The last bow shock-producing star,
[M2002]\,SMC\,83962, is located to the east from the association
\object{NGC\,602c} (Fig.\,\ref{fig:12}), which contains the only
WO star, SK\,188, known in the SMC\footnote{Actualy, SK\,188 is a
binary system composed of WO4 and O\,V stars (Foellmi, Moffat \&
Guerrero \cite{fo03a}).}. The relative position of the bow shock,
the star, and the association on the sky are consistent with the
possibility that [M2002]\,SMC\,83962 was ejected from NGC\,602c.


\begin{figure}
\includegraphics[width=9cm]{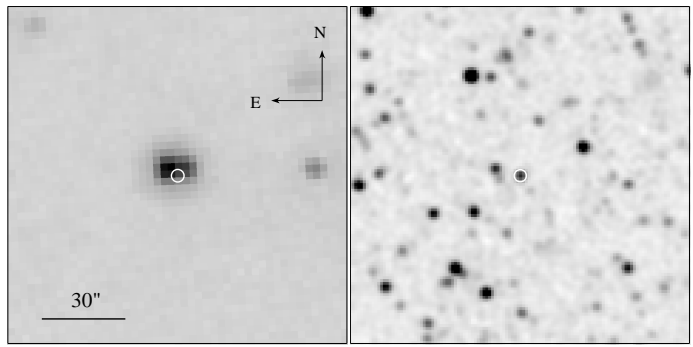}
\centering \caption{{\it Left}: MIPS $24\,\mu$m image of a nebula
attached to the WN3h+O9 star SMC-WR3 (AzV\,60a). The position of
SMC-WR3 is marked by a circle. {\it Right}: 2MASS J band image of
the same field.} \label{fig:WR}
\end{figure}

\begin{figure}
\includegraphics[width=9cm]{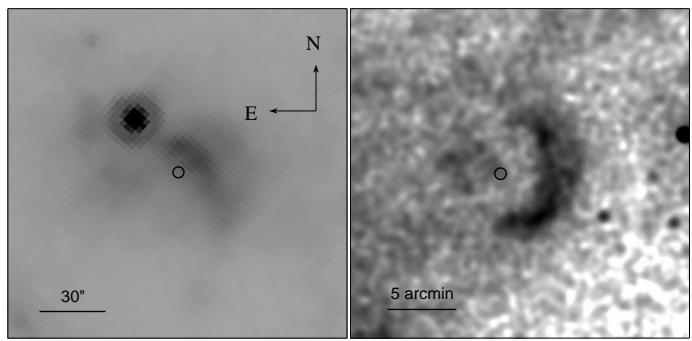}
\centering \caption{{\it Left}: MIPS $24\,\mu$m image of a nebula
around the WN2b(h) star Brey\,2 (indicated by a circle). {\it
Right}: {\it MSX} $14.7 \,\mu$m image of a nebula associated with
the WN4 star WR\,18 (indicated by a circle).}
\label{fig:Brey2+WR18}
\end{figure}

We also searched for infrared nebulae around all 12 known WR stars
in the SMC; see Massey, Olsen \& Parker (\cite{ma03}) for a list
of these stars, and Fig.\,\ref{fig:SMCbows} for their distribution
on the sky. We detected a kidney-shaped nebula attached to the
WN3h+O9 (Foellmi et al. \cite{fo03a}) star SMC-WR3 (also known as
\object{AzV\,60a}; indicated in Fig.\,\ref{fig:SMCbows} by a large
cross). The geometry of the nebula (Fig.\,\ref{fig:WR}) suggests
that it could be either a bow shock or a circumstellar nebula,
similar to those produced by the WN6 (Hamann, Gr\"{a}fener \&
Liermann \cite{ha06}) star \object{WR\,136} in the Milky Way and
the WN8h (Smith, Shara \& Moffat \cite{sm96}) star
\object{Brey\,13} in the LMC [see, respectively, Fig.\,2i and
Fig.\,2q in Gvaramadze, Kniazev \& Fabrika (\cite{gv10c}) for the
MIPS $24\,\mu$m images of these nebulae]. But because
circumstellar nebulae produced by WR stars are observed
exclusively around late-type WN stars (Gvaramadze et al.
\cite{gv10c} and references therein), that is, young WR stars
whose wind still interacts with the dense circumstellar material
(Gvaramadze et al. \cite{gv09a}), we inclined to interpret the
detected nebula as an interstellar structure, either a bow shock
created ahead of the supersonically moving star or a bright rim of
a wind-driven bubble blown-up in the inhomogeneous (dense) ambient
interstellar medium. Nebulae of similar (linear) size and
morphology were detected around the WN2b(h) (Foellmi, Moffat \&
Guerrero \cite{fo03b}) star \object{Brey\,2} in the LMC (Chu, Weis
\& Garnett \cite{ch99}; see also Fig.\,\ref{fig:Brey2+WR18}a for
the MIPS 24 $\mu$m image of the nebula) and the Galactic WN4
(Hamann et al. \cite{ha06}) star \object{WR\,18} [see
Fig.\,\ref{fig:Brey2+WR18}b for the {\it Midcourse Space
Experiment} ({\it MSX}) satellite (Price et al. \cite{pr01}) image
of the nebula]. Both nebulae have normal chemical composition
(Garnett \& Chu \cite{ga94}; Esteban et al. \cite{es93}) and
therefore are composed of a material swept-up by the stellar wind
from the local interstellar medium. We predict that the nebula
associated with SMC-WR3 should also have the same abundances as
the ambient interstellar medium. Moreover, if the detected nebula
is a bow shock, then the possible birthplace of SMC-WR3 is the
$\simeq 4$ Myr old association \object{SMC\,ASS\,16} located at
$6\farcm7$ (or $\simeq 120$ pc in projection) to the southwest of
the star.


\section{Discussion and conclusion}

The detection of bow shocks associated with a dozen field OB stars
in the SMC unambiguously proves their runaway nature. It is
therefore likely that most (and probably all) massive stars in the
field are runaways as well (cf. de Wit et al. \cite{de05};
Schilbach \& R\"{o}ser \cite{sc08}; Gvaramadze \& Bomans
\cite{gv08b}; Pflamm-Altenburg \& Kroupa \cite{pf10}). Some
support for this possibility comes from the observational fact
that the velocity dispersion for the O-type stars in the field of
the SMC is greater by $\simeq 25$ per cent than that for the O
stars located within 30 pc from the known clusters and
associations (Evans \& Howarth \cite{ev08}).

Another interesting finding by Evans \& Howarth (\cite{ev08}) is
that the velocity dispersion for {\it all} O stars in the SMC is
larger than that for the less massive (BA) stars. Evans \& Howarth
(\cite{ev08}) suggested that this difference is owing to the
effect of undetected massive binaries and runaways. This
suggestion is consistent with the observational fact that the
percentage of runaway stars is highest among the O stars (e.g.
Gies \cite{gi87}; Blaauw \cite{bl93}; Zinnecker \& Yorke
\cite{zi07}) and that the binary frequency increases with the
stellar mass (Larson \cite{la01}; Clark et al. \cite{cl08}).

Like in the Milky Way and the LMC, the runaway nature of some
field stars in the SMC can be confirmed via detection of their
high peculiar radial velocities. Measurements of radial velocities
for $\ga 2000$ stars in the SMC by Evans \& Howarth (\cite{ev08})
show that a significant number of OB stars have heliocentric
radial velocities much higher than the SMC's systemic velocity, so
that one can expect that at least some of these stars are runaways
(cf. Lamb \& Oey \cite{la08}). The sample of stars studied by
Evans \& Howarth (\cite{ev08}) contains only one of the dozen of
bow shock-producing stars listed in our Table\,\ref{tab:run}.
Interestingly, the radial velocity of this star, AzV\,471, is
offset by $\simeq 40 \, \kms$ from the local systemic velocity of
the star-forming region N83/N84, where the star is situated. This
offset along with the detection of the bow shock generated by
AzV\,471 undoubtedly proves the runaway nature of the star. Thus,
one might expect that measurements of radial velocities for other
stars from our list would result in the detection of similar
offsets as well.

In our search for bow shock-producing stars we found that some of
them are located (at least in projection) within the confines of
known associations. The presence of bow shocks around these stars,
however, implies that they are runaways and therefore were instead
injected into their host associations from the nearby ones (cf.
Gvaramadze \& Bomans \cite{gv08b}; Paper\,I). From this it follows
that the contamination of the stellar content of associations by
``alien" OB stars may contribute to the age spread observed for
some young stellar systems (e.g. Massey \cite{mas03}).

To conclude, we note that searches for bow shocks with the next
generation of space infrared telescopes along with future
high-precision proper motion measurements for massive field stars
with the space astrometry mission {\it Gaia} will allow us to
solve the problem of whether the massive stars form solely in the
clustered mode and subsequently leave their parent clusters
because of the gravitational interaction with other massive stars
(Poveda, Ruiz \& Allen \cite{po67}; Leonard \& Duncan \cite{le90};
see also Kroupa \cite{kr98}; Pflamm-Altenburg \& Kroupa
\cite{pf06}; Gvaramadze, Gualandris \& Portegies Zwart
\cite{gv09b}; Gvaramadze \& Gualandris \cite{gv10}), or whether
they can also form {\it in situ} (in the field and/or in low-mass
clusters; e.g. Oey et al. \cite{oe04}; Parker \& Goodwin
\cite{pa07}).

\begin{acknowledgements}
We are grateful to the referee for a comment allowing us to
improve the presentation of the paper. VVG acknowledges financial
support from the Deutsche Forschungsgemeinschaft. This research
has made use of the NASA/IPAC Infrared Science Archive, which is
operated by the Jet Propulsion Laboratory, California Institute of
Technology, under contract with the National Aeronautics and Space
Administration, the NASA's {\it SkyView} facility
(http://skyview.gsfc.nasa.gov) located at NASA Goddard Space
Flight Center, the SIMBAD database, and the VizieR catalogue
access tool, both operated at CDS, Strasbourg, France.
\end{acknowledgements}

\end{document}